\documentclass [floatfix, twocolumn, showpacs, aps, pra, 10pt]{revtex4-1}

\usepackage{amsmath}
\usepackage{hyperref}
\usepackage{graphicx}
\usepackage{bbm}
\usepackage{amssymb}
\usepackage[english]{babel}
\usepackage[latin1]{inputenc}
\usepackage{lmodern}
\usepackage[T1]{fontenc}

\graphicspath{{Figs/}}

\DeclareMathOperator{\e}{{\displaystyle e}}
\DeclareMathOperator{\de}{{\displaystyle d}}

\graphicspath{{Figs/}}

\usepackage{amssymb}

\begin{document}

\title{Multichannel optical atomic magnetometer operating in unshielded environment}

\author{Giuseppe Bevilacqua}
\affiliation{DIISM, University of Siena - Italy}

\author{Valerio Biancalana} 
\email{valerio.biancalana@unisi.it}
\affiliation{DIISM, University of Siena - Italy}

\author{Piero Chessa}
\affiliation{DIISM, University of Siena - Italy}

\author{Yordanka Dancheva}
\affiliation{DSFTA, University of Siena - Italy}

\begin{abstract}
A multi-channel atomic magnetometer operating in an unshielded environment is described and characterised. The magnetometer is based on D$_1$ optical pumping and D$_2$ polarimetry of Cs vapour contained in gas-buffered cells. Several technical implementations are described and discussed in detail. The demonstrated sensitivity of the setup is  $100 \, \mbox{fT}/\sqrt{\mbox{Hz}}$ when operating in the difference mode.
\end{abstract}







\date{\today}

\maketitle


\section{Introduction}
\label{Introduction}









High sensitivity magnetometers operating in geomagnetic and lower level fields  have a wide range of applications in several branches of both fundamental  and applied research \cite{pendlebury_prd_15, budker_natph_07, marmugi_ol_14}. They are used   to both measure and track large scale field variations, such as in geomagnetic surveys \cite{lenci_jpb_12}, and to detect local magnetic variations, such as in measurements of ultra-low-field NMR \cite{savukov_prl_05, biancalana_arnmrs_13,  bevilacqua_jmr_16} and  MRI \cite{savukov_jmr_13}, in relaxometry \cite{ganssle_2014, dolgovskiy_jmmm_15}, in biomagnetic applications such as magneto-cardiography \cite{belfi_josab_07, bison_apl_09, kamada_pm_12, alem_pmb_15} and in magneto-encephalography \cite{johnson_apl_10, romalis_2006}. The detection of weak local signals can be performed in small shielded volumes, and indeed this approach is preferable when small sized sources need to be characterised. However, the existence of large sized samples (such as human bodies in biomagnetic measurements), the high costs and the delicateness of shielding devices may prompt the use of alternative methods in order to reject the magnetic noise generated by far-located sources. 

Multi-channel sensors \cite{bison_apl_09, alem_pmb_15} open the way for the active compensation of common mode magnetic signals as well as the direct measurement of difference mode signals. Thus, even if top-level sensitivity normally requires  operation in shielded volumes, relatively high-sensitivity magnetometry can be profitably performed in open volumes \cite{bevilacqua_jmr_09, bevilacqua_jmr_16} with a more simplified implementation  when using sensor operating in non-vanishing fields. 
Superconducting quantum interference devices (SQUIDs) have traditionally provided the highest magnetometric sensitivity. Recently, the most sensitive atomic magnetometers in the spin-exchange-relaxation-free (SERF) regime, which operate in vanishing fields, have demonstrated their competitiveness with SQUIDs, while also having the advantage of not requiring cryogenic cooling, and of being unaffected by electric noise. Compared to SQUIDs and SERF, other atomic magnetometers have worse sensitivity, although their performance remains better than $1~\mbox{pT}/\sqrt{\mbox{Hz}}$, meaning that they are unrivalled by other simpler devices such as Hall and magnetoresistance sensors. 

Atomic magnetometry is based on the detection of the Larmor precession of an optically pumped vapour in the field under measurement. 
The basic working principles of atomic magnetometers date back to over a century ago \cite{macaluso_98, macaluso_99}, and the first pioneering implementations date back several decades \cite{bell_prl_61}. Nevertheless,  renewed interest in atomic magnetometry has arisen much more recently, thanks to progresses in the knowledge of optical pumping processes \cite{happer_rmp_72, auzinsh_2010}, and to technical achievements regarding laser sources and atomic sample manufacturing \cite{shah_nat_07}. This set of circumstances has led to important advances in the last decade, so that  atomic magnetometers now represent a versatile class of sensors with state-of-the-art sensitivity \cite{budker_natph_07, chalupczak_apl_12, romalis_prl_13}. 
In contrast to SQUID magnetometers, which are vector component detectors that measure the field projection along one particular direction, atomic magnetometers (at least in their basic configuration) are scalar in nature, as they measure the modulus of the magnetic field. A scalar sensor operating in a dominant bias field is mainly sensitive  to the field variation along the bias field. Thus atomic magnetometers operating in non-vanishing fields are responsive to one component. 

Provided that the environmental (bias) field is relatively homogeneous, this kind of magnetometers can be profitably organised into sets of two or more elements to produce a gradiometric response \cite{belfi_josa_09, bison_apl_09}, thus rejecting common-mode magnetic noise and identifying signals as weak as several orders of magnitudes below the bias field.

The literature reports several kinds of setups, where various parameters of the pumping radiation are modulated, such as the polarisation \cite{breschi_apl_14}, the intensity \cite{junhai_cpl_12} or the optical frequency \cite{belfi_josa_09}. A comparison of the different approaches is presented in Ref.\cite{grujic_pra_13}. The frequency modulation of the pumping radiation simplifies the implementation (no optical modulators needed), particularly in a dual radiation setup.

In this paper we describe an all-optical atomic magnetometer based on Cs vapour, optically pumped by means of a frequency modulated laser source tuned to the D$_1$ line, and probed through the rotation of the polarisation plane of a weak linearly polarised beam tuned to the D$_2$ line. 
The system contains two composite sensors and the output signals are extracted from two couples of half-beam polarimeters, thus obtaining a four-channel system. Depending on the  channels selected and on  choices at the data analysis stage, this setup enables gradiometric measurements of various orders and with several differently arranged baselines.
The  sensitivity demonstrated is limited by  residual magnetic noise at low frequencies, and, when using the shortest baseline, is essentially determined by the instrumental noise. In this latter configuration the sensitivity level ($100~\mbox{fT}/\sqrt{\mbox{Hz}}$ in the range  20~Hz$\div$100~Hz) is comparable to values reported in the literature for similar setups operating in a shielded environment.

\section{Setup overview}
\label{sec:setup}
\subsection {Optics}
\label{subsec:optics}
\begin{figure}[htbp]
   \centering
  \includegraphics [width=\columnwidth] {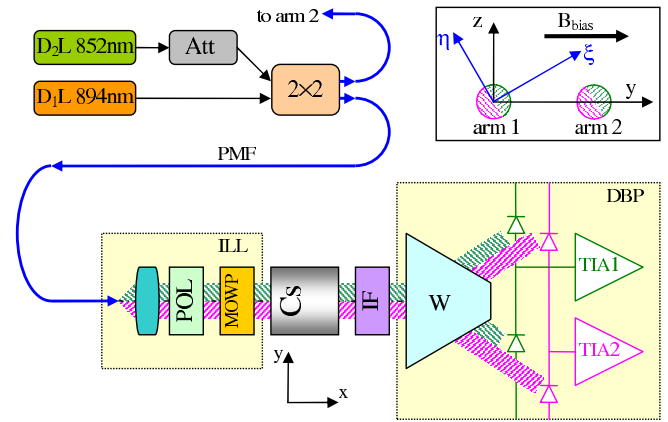}
  \caption{Schematics of  a single arm of  the magnetometer. The
    D$_1$L  pumps the  atoms and  the D$_2$L  probes them.  The two  radiations  are  mixed in  the
    $2\times2$  coupler, and  led  to the  sensor  via a  polarisation-maintaining fibre (PMF). At the PMF output, the illuminator (ILL) collimates the two radiations into a single beam and adjusts    their polarisation by means of a  multi-order
    wave-plate (MOWP), which renders  the polarisation of the pump circular, and leaves  the polarisation of the probe linear. The Cs 
    cell  is   illuminated  by  the two
    co-propagating  radiations.    After  interaction with the Cs vapour, the pump radiation is stopped
    by  an  interference  filter  (IF), while the probe impinges on a Wollaston polariser (W) oriented at $45^\circ$ with respect to the input polarisation plane. A dual balanced  polarimeter (DBP)
    measures the rotation of the probe polarisation plane  for the two halves of the beam separately. To this end, the  imbalances in the photocurrents flowing in the two couples of series silicon photodiodes are amplified by trans-impedance amplifiers (TIA). The inset represents  a front view of the probe beams of both the arms, where the relative orientations of  the various baselines and of the bias magnetic field are defined.}
  \label{fig:setup}
\end{figure}

This magnetometer is the evolution of an apparatus described previously \cite{belfi_josa_09, biancalana_arnmrs_13}. It has two identical arms for differential measurements,
but also  allows   for   quadruple   channel  operation.   Each  arm   (see
Fig.~\ref{fig:setup}) contains  an illuminator (ILL) providing  two co-propagating radiations, a Cs vapour cell, and a dual balanced polarimeter. One of the radiations is resonant
 with the D$_1$ transitions of  the Cs atoms  and is used to optically pump the
atoms; the other radiation is near resonant with the D$_2$  transitions and is used to  probe the atomic
spins precession. The
light generated by the pump laser  D$_1$L and by the probe laser D$_2$L is
coupled to  a polarisation-maintaining fibre (PMF).  A  proper mixture
of the two lights  is accomplished using a
$2\times  2$ PM  on-fibre mixer. Both  light sources  are based  on solid-state
continuous-wave lasers. The probe  radiation is attenuated by means of
an  on-fibre attenuator (Att) and is kept at a constant frequency, while the pump  radiation is  left at  the mW level, and is broadly frequency modulated.   In the proximity of  each
vapour  cell, the  two  radiations  are
collimated  into  a beam  of about  10~mm in  diameter, their
linear polarisation is reinforced by a polarising cube, and finally a
special  multi-order waveplate (MOWP)  makes  the pump radiation
circularly polarised, while leaving the probe one linearly polarised. 

For this purpose we took advantage of quartz's property of birefringence at the D$_1$ and D$_2$ transition wavelengths in Cs. In fact, it is possible \cite{optisource_priv_10} to design a multi-order quartz waveplate with a nominal relative delay of 4.75 wavelengths for  D$_1$ radiation and 5.00045 wavelengths for  D$_2$ radiation. Thanks to their low  order, the relative delay of such waveplates is only weakly dependent upon misalignment (less than 0.1 \% for $1\deg$ misalignment).

The MOWP technique was similarly employed in another setup described in the literature \cite{johnson_apl_10}, with several practical advantages that are worthy of mention. In fact, the use of different transitions for the pump and probe processes, facilitates both the alignment (the two radiations are mixed in one fibre, and propagate after collimation  along a single optical axis) and the pump suppression after the interaction. This is effectively accomplished by means of an interference filter.

\subsection{Sensor}
\label{subsec:sensor}
The  atomic  sensors  are  made  of high quality sealed  cells with flat windows, containing    solid source Cs and
23~torr  of N$_2$  as a  buffer gas.   The  vapour density of the Cs is increased by
warming  the  cells up  to  about  $45^\circ$C by means of an  alternating
current heater. The heater is made of two coaxial coils of thin copper wire wound in an anti-inductive configuration. Possible spurious  magnetic fields generated by the heater are rendered of negligible relevance \cite{bevilacqua_pra_12} by supplying the coils at $50$~kHz, a value largely exceeding the atomic Larmor frequency.

After  interaction  with the  atomic  vapour,  the  pump beam  is
blocked  by an  interference  filter  (IF) [Thorlabs 850/40, with a transmission of 98.5\% at 852~nm and an optical density of 4.2 at 894~nm] and   the  probe  beam
polarisation is analysed by dual balanced polarimeters (DBP) made of a
45$^\circ$  oriented Wollaston  analyser  (W) and  two  couples of  silicon
photo-detectors. A  similar technique, based on segmented photodetectors, has  recently been applied in a higher buffer-gas pressure potassium setup \cite{keder_aip_14}. 
Each  couple of photodetectors measures the polarisation of half a beam (a semicircle in the inset of Fig.~\ref{fig:setup}) by converting the polarisation rotation into a photo-current imbalance, which is in turn amplified and converted into voltage by 
low-noise  trans-impedance  amplifiers  (TIA).  
The signal extracted by a balanced polarimeter is proportional to the sinus of twice the polarisation rotation angle. At the maximal observed rotation (about 0.15 rad) the linear approximation results in a 0.35\% underestimation, which in the data elaboration is neglected.

The operator may select the channels to be acquired and the corresponding TIA outputs are simultaneously digitised by means of a 16 bit  data acquisition card. 

The sensors work in a  homogeneous magnetic field $\vec B_{\mathrm {bias}}$, which is obtained by partially compensating the environmental one and is oriented along the $y$ axis. The longitudinal field component ($x$) is zeroed, and thus the sample magnetisation, which is periodically reinforced along the optical axis, precesses in the $xz$ plane. Small variations in the field slightly change the resonant precession frequency, or (as discussed in the Appendix) the precession phase when periodic resonant pumping is applied.

\section{Principle of operation}
\label{sec:principle}
\subsection{Laser sources and modulation of the optical frequency}

The  probe  radiation  is  generated by  D$_2$L, a  single-mode,  Fabry-Perot,
pigtailed  diode  laser  that is near-resonant  with  the  D$_2$  line  of  Cs  at
852~nm.  This  illuminates the  cell  at a  weak intensity,  which is 
attenuated  down to  a level  of  a  few  $\mu\mbox{W/cm}^2$ in  order  to
negligibly perturb the  atomic state measured.

The  pump  radiation  is   generated  by D$_1$L,  a  single-mode,  distributed
feedback, pigtailed diode laser. Its optical frequency is controlled
through the junction current and is periodically made resonant to the
transition      $|^2S_{1/2}, F_g=3  \rangle \rightarrow
|^2P_{1/2} \rangle$ of  Cs  at  894~nm. The  optical frequency of the pump  laser is modulated
using  a square wave  signal generated  by a computer-controlled digital-to-analogue converter.

\begin{figure}
\includegraphics [width=\columnwidth] {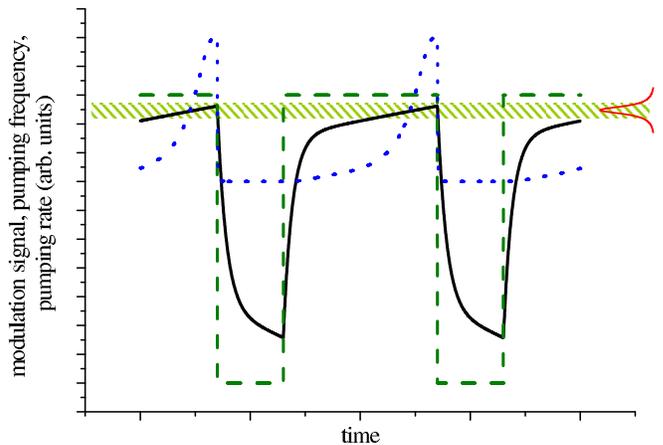}
\centering
\caption{Optical frequency of the D$_1$L (solid black) in response to a square wave modulation signal (dashed green). The green hatched band points out the time intervals when D$_1$L goes into resonance with an atomic line (rightmost red profile), and the blue dotted line represents the resulting pumping rate. The figure is not to scale, but is produced in accordance with the simple model described by eq.~\ref{eq:modulazmodel}. In the text the maximal and minimal frequencies reached in the period will be referred to as \textit{front}- and \textit{back}-frequency, respectively.}
\label{fig:modulaz}
\end{figure}

Applying a square wave modulation to the junction current of D$_1$L produces a modulation of the optical frequency that cannot  easily be described in terms of frequency steps. At least three aspects have to be taken into account in order to evaluate the response to sudden changes in the modulating signal. First of all, the modulation input of the laser driver has a limited bandwidth. This acts as a low-pass filter with a cut-off frequency set at about $\Gamma_1=2\pi \times$25kHz in our case. In addition, when changing the junction current, the laser emission changes for two reasons. The first is related to the change of  refraction index and has a very quick (instantaneous) response. The second reason is a thermal change of the cavity length, and is a much slower process. The latter introduces a second cut-off frequency at $\Gamma_2 \approx 2\pi \times$30~Hz.
We performed a detailed characterisation, based on techniques involving both atomic and interferometric references  \cite{dancheva_rsi_13}). The optical frequency response to a signal at $\omega$ can be modelled as the sum of two low-pass-filtered terms, and the instantaneous optical frequency deviation can be approximated as
\begin{equation}
\Delta \nu=  \sum_{n=0}^{+\infty} 
\left (\frac{w_1 \Gamma_1}{\Gamma_1+i n\omega } +
\frac{w_2 \Gamma_2}{\Gamma_2+ i n\omega_0 } \right )
V_n  + c.c.
\label{eq:modulazmodel}
\end{equation}
where $\Gamma_i$ and $w_i$ are the cut-off frequencies and weights of the thermal and input filters, respectively, and $V_n$ are the square wave Fourier components. More precisely, the modulation signal
\begin{equation}
V(t)=\sum_{n=0}^{+\infty}V_n + c.c.=\sum_{n=0}^{+\infty}\frac{2A}{n \pi} \sin (n \pi \delta) 
e^ {n i \omega t} + c.c.
\end{equation}
is a square wave with amplitude  $A$  and duty cycle $\delta$. Typically, the conditions, $\Gamma_1 \ll \omega$ and $\Gamma_2 > \omega$ hold. 
The resulting optical frequency modulation is represented in Fig.~\ref{fig:modulaz}.

\subsection{Atom-light interaction}
The D$_1$L optical frequency deviation amounts to tens of GHz, thus the radiation periodically goes in and out of resonance with given transitions of the D$_1$ line.

\begin{figure}
\includegraphics [width=\columnwidth] {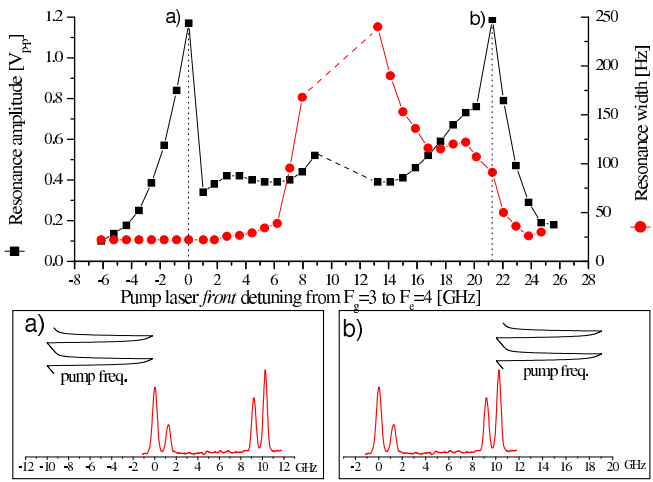}
\centering
\caption{Amplitude and width of the observed resonance line  for a fixed D$_1$L modulation span, as a function of the front frequency. A square wave modulation signal with a duty cycle of $75\%$ is applied. This  produces (see  Fig.~\ref{fig:modulaz} and eq.~\ref{eq:modulazmodel}) the instantaneous D$_1$L frequency represented by black lines in the insets, having a span of 11.2~GHz. Two amplitude maxima are observed when either (a) the front frequency is resonant with the leftmost hyperfine component or (b) the back frequency is resonant with the rightmost one. In  case (a) the resonance width is minimum, which makes this configuration the optimal working condition. The dashed lines correspond to intermediate detunings, at which no clear resonance line is observed.
}
\label{fig:opt_pump}
\end{figure}

Let us define the highest instantaneous optical frequency as \textit{front} frequency and correspondingly the lowest one as \textit{back} frequency (see Figs.~\ref{fig:modulaz} and \ref{fig:opt_pump}).  If the frequency deviation is kept constant,  when the \textit{front} and the \textit{back} frequencies increase, the amplitude and the width of the magnetic resonance change, as shown in Fig.~\ref{fig:opt_pump}. The probe laser is cw and tuned to be nearly resonant with the triplet $| ^2S_{1/2}, F_g=4 \rangle \rightarrow | ^2P_{3/2} \rangle$, thus it probes the atomic precession of the $F_g=4$ state. When the D$_1$L approaches the D$_1$ line from the side of the $|^2S_{1/2}, F_g=3 \rangle \rightarrow | ^2P_{1/2}, F_e=4 \rangle$ transition, the amplitude of the resonance increases, while its width remains almost constant. The amplitude shows two maxima occurring when the \textit{front} frequency goes into resonance with the transitions 
$| ^2S_{1/2}, F_g=3 \rangle \rightarrow | ^2P_{1/2}, F_e=3,4 \rangle $ (see Fig.~\ref{fig:opt_pump}a)) and when the  \textit{back} frequency goes into resonance with the $|^2S_{1/2}, F_g=4 \rightarrow | ^2P_{1/2} \rangle$ (see Fig.~\ref{fig:opt_pump}b)), respectively.

When D$_1$L is near-resonant or resonant with the $F_g=4$ (either with its \textit{front} or \textit{back} frequency),  the resonance width increases dramatically untile the resonance is barely recognisable. In the rightmost region, when the back frequency is near-resonant with the $F_g=4$ (Fig.~\ref{fig:opt_pump}b),  the resonance amplitude grows again (high-rate Zeeman optical pumping is performed), but the resonance is much broader. Moreover, in this condition a small red detuning results in an abrupt decrease in the signal amplitude, due to hyperfine optical pumping toward the unprobed $F_g=3$ ground state.

\subsection{Sensitivity and optimal working conditions}
The modulation duty cycle, amplitude and offset are adjusted to ensure
maximum sensitivity. As seen from Fig.~\ref{fig:opt_pump}, the best sensitivity is achieved when  the D$_1$L \textit{front} frequency  is  resonant with the $|^2S_{1/2}, F_g=3  \rangle \rightarrow |^2P_{1/2}\rangle$ hyperfine components  of the D$_1$ line. Concerning the probe laser, its optimal frequency is found to be 2~GHz blue detuned with respect to the triplet  $|^2S_{1/2},  F_g=4  \rangle \rightarrow  |^2P_{3/2}\rangle$, which  is an acceptable compromise in order to achieve a good polarisation rotation signal and small excitation rate.

Under these conditions, the pump laser plays two roles: it causes a strong hyperfine optical pumping of the Cs atoms towards the  ground state analysed, and simultaneously pumps the atoms into a specific Zeeman sublevel of that state, due to its circular polarisation.   
The combination of these two effects ensures a high amplitude and small linewidth of the resonance, also thanks to the effect known as light narrowing \cite{appelt_pra_99, scholtes_pra_11}. The linewidth is also narrow  thanks to the fact that the  hyperfine ground state measured is weakly perturbed, because D$_1$L is  either far off-resonance or resonant with the other (depleted)  hyperfine  ground  state.

The Zeeman  pumping produces magnetisation of the Cs vapour along the wave-vector ($x$ direction). As the magnetic  field is transverse to the pump light propagation ($y$ direction), the   atomic state where atoms are pumped evolves  in  time. At a macroscopic scale this corresponds to  a magnetisation  precession in the  $xz$ plane at an angular frequency $\omega_L$, which is set by the magnetic field modulus $B$ and the gyromagnetic factor $\gamma$: $\omega_L = \gamma B$.  Clearly detectable precessing magnetisation is obtained, again at the macroscopic level, when the pump radiation interacts with the atoms synchronously with the precession.

\section{Operation modes, field range, resonance linewidth}
\label{sec:operation}

The atomic precession is detected by analysing the rotation of the polarisation plane of
a weak linearly  polarised  probe  beam,  propagating  in  the  $x$
direction.  As is known, the linear polarisation can be regarded as a superposition of two,  $\sigma^+$ and $\sigma^-$, counter-rotating circular polarisation  components, which  experience  opposite variations of the refraction index (time dependent circular birifringence). Along the propagation, these two components accumulate different phase delays, and superpose into a linear polarisation rotated by an angle oscillating  at frequency set by the atomic precession.   Beside the different dispersions,  the  $\sigma^+$ and  $\sigma^-$  components experience  also different levels of absorption (circular dichroism),  but the relevance of this  aspect is  limited by appropriately detuning the probe  laser from the  centre of the atomic resonance.

The polarimetric signal can be analysed as a function of the D$_1$L modulation frequency in order to characterise  the   atomic  magnetic
resonance, in an operation mode that is referred to as \textit{scanning mode} and enables accurate evaluation of both  the precession frequency
(used to adjust $B_{\mathrm{bias}}$) and  the  resonance linewidth observed. The latter is the parameter to be minimised (see below) in the procedure devoted to reducing the field inhomogeneities.

For short-time measurements, it is instead possible to modulate D$_1$L with a periodic signal near-resonant with the Larmor precession, and to infer the magnetic field from the dephasing of the polarimetric signal. This operation mode is referred to as \textit{free-running}. It unavoidably suffers from slow drifts of the environmental field, which can eventually lead the system too far from the resonance. This problem is counteracted in a third operation mode (\textit{locked mode}). In this case one of the polarimetric signals is used as an input   for  a
phase-locked loop (PLL) to stabilise the magnetic field \cite{belfi_rsi_10}. The environmental noise is reduced within the loop bandwidth, and this improves the noise rejection when performing differential measurements, as the spurious common mode noise terms are reduced.

\begin{figure}
\includegraphics [width=\columnwidth] {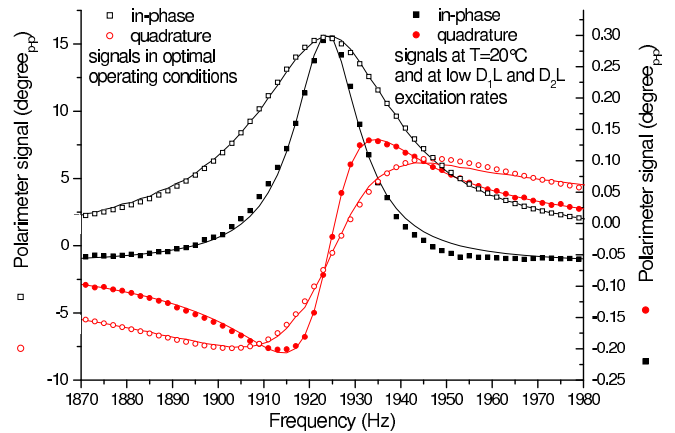}
\centering
\caption{Polarisation rotation of D$_2$L radiation (in-phase and quadrature components)   as  a   function  of   the  D$_1$L  modulation    frequency. Dots: values measured, line: best-fit curve, modelled with a Lorentzian profile. The best fit estimation of HWHM is 20~Hz under the optimal working conditions, in which the maximum  rotation of $16^\circ$ peak-to-peak is observed. A narrower line (10~Hz) is obtained at lower temperatures and lower excitation rates, at the expenses of a much smaller rotation ($0.3^\circ$).}
\label{fig:line}
\end{figure}

The magnetometer  operates  in  bias  magnetic   fields  ranging  from
$100~\mbox{nT}$ to $6~\mu\mbox{T}$.
Fig.~\ref{fig:line}  shows the  in-phase  and quadrature
signals of the Cs magnetic resonance as functions of  the  pump laser  modulation frequency, as recorded in the \textit{scanning mode}.  The
 linewidth measured (half-width at half  maximum, HWHM) is of the order
of 20~Hz and  is determined by the spin-exchange  collisions and power
broadening contributions. 

As shown in the Appendix, this linewidth sets the response bandwidth of this kind of sensors, which --with the exception of some specific implementations \cite{terao_pra_13}-- corresponds to the cut-off frequency of their response.

The direction of the bias magnetic field is set  with the help  of three
pairs    of     1.8~m    side    square     Helmholtz    coils    (see
Refs.  \cite{belfi_josab_07,  belfi_rsi_10}  for additional  details).
The magnetic field Jacobian,  $G_{ij}=\partial B_i / \partial x_j$, is
zeroed  using five quadrupole devices, taking  specific  care with  the
$\partial  B_y  / \partial  x_i$  elements,  which cause static inhomogeneities of $\delta B_{\lVert}$ and thus most affect  the performance.  This  is coarsely achieved  with permanent magnets  and then finely  adjusted with electromagnets. After such field inhomogeneity compensation, the resonance  narrowing  is   limited  --under  operative conditions-- by the spin-exchange collisions and pump and  probe excitation rates. As a matter of fact, when operating at very low Cs densities and excitation rates (at about $10^\circ$~C), resonances as narrow as $6$~Hz are recorded.

In the \textit{free-running} and \textit{locked} modes the field variations are inferred from phase estimations of the polarimetric signal. As shown in the Appendix, the phase $\varphi (t)$ responds to the time-dependent field $\delta B_{\lVert}(t)$, according to a first order equation (see eq. \eqref{eq:fin:phi:lin}),
so that working at the centre of resonance ($\Delta=0$) its variation is promptly inferred from the measured $\varphi$ as
\begin{equation}
\delta B_i=\gamma^{-1}\left [\Gamma \varphi_i+ 2 f_S \left (\varphi_{i+1}-\varphi_{i-1} \right) \right ]
\end{equation}
 $f_S$ being the sampling frequency,   $\delta B_i$ the field variation and $\varphi_i$ the phase, sampled at the times {$t_i$}.

\section{Noise and higher-order gradiometry}
\label{sec:noise}
As far located (noise)  field  sources  produce mainly common-mode magnetic signals,  their contribution can be effectively rejected by performing differential measurements.
The principal arms have an adjustable base-line $5.6~\mbox{cm}$ in the $y$ direction, with a minimum of $5.6~\mbox{cm}$.   Each arm is in turn divided in two secondary sections by using segmented photodiodes, as discussed in Sec~\ref{subsec:sensor}. The array elements are arranged in such a way as to measure the two halves of the laser spot independently, which enables gradiometric measurements  with  a fixed baseline   of  $0.5~\mbox{cm}$, as well as higher-order gradiometric measurements.

The N$_2$ buffer gas contained in the sensor cells quenches  the  fluorescence light avoiding radiation trapping, and slows down  the thermal  motion of Cs  atoms, making it  diffusive.  The diffusion constant of Cs in N$_2$ is about D=2.4~cm$^{2}$/s at 23~Torr \cite{beverini_pra_71}.
Over a time interval as long as the spin relaxation time ($1/2\pi\Gamma\approx 8$~ms), diffusive displacements of the order of $\Delta x =$1.4~mm are thus expected to occur: a distance definitely shorter than the beam radius. The short baseline operation is thus  responsive to field inhomogeneities.

If the illuminator and the dual polarimeter (ILL and DBP, respectively, in Fig.~\ref{fig:setup}) of an arm are rigidly mounted,  so that they rotate synchronously around the optical axis, gradiometric detection is performed along an arbitrary direction $\xi$ of the $yz$ plane (see the inset in Fig.~\ref{fig:setup}) and the system  responds with the short baseline to $\partial B_y / \partial \xi$. 
The long baseline is oriented along the $y$ direction,  thus the four channel operation enables  measurements of second order terms  $\partial ^2 B_y / \partial \xi \partial y$, which (depending on the alignment) can be varied from $\partial ^2 B_y / \partial y^2$ to $\partial ^2 B_y / \partial y \partial z$.

\begin{figure}
\includegraphics [width=\columnwidth] {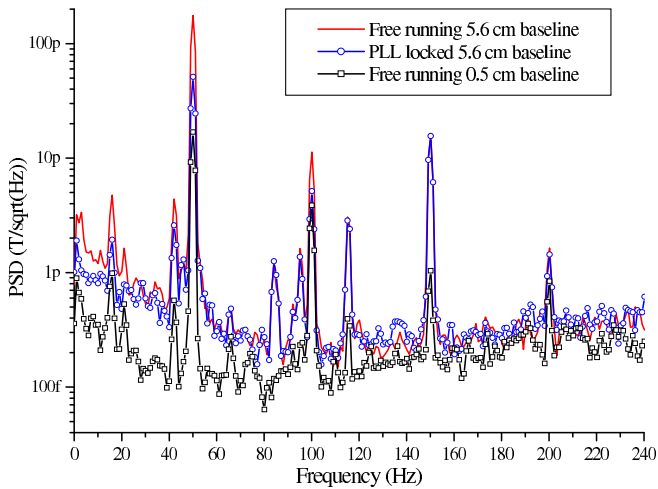}
\centering
\caption{Noise patterns of the difference between two signals obtained in two operation  modes. The  traces show the RMS values of the power spectral density averaged  over 10 measurements. The three spectra correspond to the operation modes reported in the inset.  }
\label{fig:noise}
\end{figure}

Hereafter, let us consider the case in which the three baselines are parallel, i.e. each couple of channels responds proportionally to $\partial B_y / \partial y$. 
When combining the signals from two channels different   couples   can be chosen,  and correspondingly different   residual   noise  is
observed. As expected, the bigger the baseline the higher is the residual
noise, as  shown in Fig.~\ref{fig:noise}.   

When applying the PLL field-stabilisation system \cite{belfi_rsi_10} the noise in
the $y$ direction  is actively compensated, which decreases the  noise  in the  frequency  range  from 0  to  200~Hz,  that is  the
bandwidth  of  the  PLL.  The  ultimate  estimated  magnetometer
sensitivity, which results from the superposition of detection noise and residual magnetic noise, is of the order of $100$~fT/$\sqrt{\mbox{Hz}}$, as can be seen in Fig.~\ref{fig:noise} from  measurements conducted over a 0.5~cm  baseline.

More generally, instead of recording the difference between two signals, it is possible to combine  three or four signals in order to achieve optimal  noise cancellation.
This task is performed by determining the coefficients of the linear superposition that minimise the variance of the combination in the absence of a signal source. In formula, when acquiring N channels, the N signals $s_k(t)$ with k=0, 1,..., N-1 are linearly combined in $S(t)=\sum\limits_{k=1}^{N-1} a_k s_k(t)$, where $a_0$ is set to 1, and $a_k$ (k=1, ..., N-1) are determined as the ones minimising $\lVert S(t) \rVert^2=\sum\limits_{i=1}^n S(t_i)^2$, $\{t_i\}$ being the sampling times. Depending on the application, the optimal coefficients $\{ a_k \}$ can be determined after having band-pass-filtered the signals $s_k(t)$, in order to achieve the best noise reduction in a specific spectral range.


\section{Conclusion}
We have developed and characterised a multichannel optical atomic magnetometer based on a dual arm setup and segmented detectors, enabling polarimetric measurements in transverse subsections of the probe beam, for a total of  four independent channels and various alignment-dependent gradiometric arrangements. The  setup described makes it possible to perform differential measurements and gradiometric detection over various baselines and different orders. The environmental (common mode) noise-rejection permits the achievement of a noise level as low as a 100~fT/$\sqrt{\mbox{Hz}}$ .

\appendix
\section{Polarimetric signal phase}
\label{sec:appendix}
This appendix is devoted to explaining the conditions and the consequent approximations considered in order to infer time-dependent magnetic field variations from the phase of the polarimetric signal.

We refer to the geometry of the experiment reported in Fig.~\ref{fig:setup}
and we assume that the magnetic field is a superposition of two parallel terms: a dominant static term (bias) and a small periodic one
\[
\mathbf{B}(t)  =   (B_{\mathrm {bias}}  +  \delta B_{\lVert}(t)) \hat{y}, 
\]
with $\delta B_{\lVert}(t+ 2\pi/\Omega) = \delta B_{\lVert}(t)$.


\noindent
 
The Larmor equations for the magnetization $\mathbf{M} =(M_x, M_y, M_z)$ are 
\begin{eqnarray*}
  \dot{M}_x & = & -\Gamma M_x + (\omega_L + \omega_1(t)) M_z + f(t)\\
  \dot{M}_y & = & - \Gamma M_y \\
  \dot{M}_z & = & -\Gamma M_z - (\omega_L + \omega_1(t)) \, M_x , 
\end{eqnarray*}
where    $\omega_L =  \gamma B_{bias}$
and  $\omega_1(t)  =  \gamma   \delta B_{\lVert}(t)$  are the Larmor frequencies corresponding to the bias field   and to its variation, respectively. In our  case $\omega_L$ is in the kHz range while the damping
$\Gamma_{\parallel}$ and  $\Gamma_{\perp}$ are in the 10 Hz range  so we
assume   they  are  equal   $\Gamma_{\parallel}  =   \Gamma_{\perp}  =
\Gamma$.  
The forcing term $f(t)$ is due to the laser optical pumping and can be
calculated exactly, in the approximation of weak laser power (details will be provided in a forthcoming paper). This term can be expanded in Fourier space as 
\[ 
f(t) = \sum_{-\infty}^{+\infty} f_n \e^{i\,n \omega t}
\]
where  $\omega$ is  the laser  modulation frequency.  The coefficients
$f_n$ satisfy $f_{-n} = f_n^*$ and  for odd $n$ it can be shown 
that $f_{n} = i |f_n|$. 

Introducing  $M_+ =  M_z  + i  M_y$  the relevant  equation of  motion
becomes 
\begin{equation}
  \label{eq:larmor}
  \dot{M}_+ = \left[  - \Gamma - i ( \omega_L  + \omega_1(t) ) \right]
  M_+ + f(t).  
\end{equation}
The steady-state ($\Gamma t \gg 1$) solution can be written as
\begin{equation}
  \label{eq:sol:ss}
  M_+(t) = \e^{-(\Gamma + i \omega_L) t - i \theta_1(t)}
  \int_0^t \e^{+(\Gamma + i \omega_L) ' + i \theta_1(t')} \, f(t')\de t' ,
\end{equation}
where $\theta_1(t)  = \int_0^t  \omega_1(t') \de t'$.  Introducing the
Fourier expansion 
\begin{equation}
  \label{eq:fourier:theta1}
  \e^{i \theta_1 (t) } = \sum_{-\infty}^{+\infty} G_n \e^{i n \Omega t}
\end{equation}
(in the case that  $B_1(t)$ is  sinusoidal  the  $G_n$  are the  usual  Bessel
functions) following some algebra we find 
\begin{equation}
  \label{eq:sol:M+:fin}
  M_+(t) = \sum_{s,n,m=-\infty}^{+\infty} \frac{G_{n-s}^* G_n f_m}{\Gamma
    + i\omega_L + i  m\omega + i n \Omega} \e^{i s  \Omega t } \e^{i m
    \omega t}. 
\end{equation}
In the  experiment the modulation frequency $\omega$  is resonant with
$\omega_L$ so that  the term with $m=-1$ is  the dominant one. Finally
we obtain the expression 
\begin{equation}
\begin{split}
  \label{eq:M+:expr}
   M_+(t) \approx \sum_{s,n=-\infty}^{+\infty} \frac{G_{n-s}^* G_n f_{-1}}{\Gamma
    + i(\omega_L -  \omega) + i n  \Omega} \e^{i s \Omega t  } \e^{- i
    \omega t} \\
    \equiv A(t) \e^{-i \omega t}, 
\end{split}
\end{equation}
which shows, as expected, that the magnetization follows the forcing term
with  a complex  amplitude $A(t)$,  which has the  periodicity of  the $\delta B_{\lVert}(t)$
field. 

Although  this   expression  is  exact,  the  double   sum  hides  the
interpretation. Moreover,  the experiment  monitors the phase of 
$ 
M_x(t) = \Re\left[ A(t)\e^{-i\omega t} \right]
$
so it  is better to work  out an equation for  $A(t)$ directly. Assume
that the solution $M_+(t)$ has the form outlined in \eqref{eq:M+:expr}
and that $A(t) = a(t) \e^{i  (\pi/2 + \varphi(t) )}$. The phase $\varphi(t)$
(the excess of  phase from exact resonance) is the quantity monitored
in the experiment. Inserting this ansatz in eq.\eqref{eq:larmor}, after some algebra we find 
\begin{subequations}
  \begin{eqnarray}
    \label{eq:a:phi}
    \dot{a} & = & - \Gamma a + \left| f_{-1} \right| \cos \varphi \\
    a \dot{\varphi} & = & - ( \Delta + \omega_1) a + \left| f_{-1} \right| \sin \varphi ,
  \end{eqnarray}
\end{subequations}
where we have introduced the detuning $\Delta\equiv \omega_L - \omega$ and
coherently  approximated $f(t)  \approx  f_{-1} \e^{-  i  \omega t}  =
|f_{-1}| \e^{i \pi/2} \e^{-i \omega t}$.  The
laser operates  at low power, so  the modulus of  $|f_{-1}| \ll \Gamma$
and the $a(t)$ adiabatically follows 
\[ 
a(t) \approx \frac{ |f_{-1}| }{\Gamma} \cos \varphi ,
\] 
giving the dynamical equation for the phase 
\begin{equation}
  \label{eq:fin:phi}
  \dot{\varphi} = - \Delta - \omega_1(t) - \Gamma \tan \varphi .
\end{equation}
This, under our experimental conditions can be linearized as
\begin{equation}
  \label{eq:fin:phi:lin}
  \dot{\varphi} = - \Delta - \omega_1(t) - \Gamma \varphi, 
\end{equation}
which describes the evolution of the phase as the sum of a damping term
($-\Gamma \varphi$)
and a forcing term ($ -\Delta - \omega_1(t) $). 

The steady-state ($\Gamma t\gg 1$) solution is now elementary 
\begin{equation}
  \label{eq:phi:fin:lin:sol}
  \varphi(t)  = \frac{\Delta}{\Gamma} +  \int_0^t \e^{-\Gamma  ( t  - t')
  }\omega_1(t') \de t' .
\end{equation}
For a sinusoidal signal $ \omega_1(t) = w \cos (\Omega t + \phi) $ one
finds 
\begin{equation}
  \label{eq:sol:phi:lin:sin}
  \varphi(t)   =   \frac{\Delta}{\Gamma}   +  \frac{w}{\sqrt{\Omega^2   +
      \Gamma^2}}
  \cos ( \Omega t + \phi - \arctan(\Omega/\Gamma) ).
\end{equation}

\acknowledgments
The  authors acknowledge  the valuable  technical support  of Leonardo
Stiaccini and the financial support of the Italian Ministry for Research
(FIRB  project  RBAP11ZJFA005). The authors thank  Emma Thorley of Language Box (Siena) for revising the English in the manuscript.

\bibliographystyle{ieeetr}
\bibliography{unsh_refs}

\end{document}